\documentclass[pdflatex,sn-mathphys-num]{sn-jnl}

\usepackage{graphicx}%
\usepackage{multirow}%
\usepackage{amsmath,amssymb,amsfonts}%
\usepackage{amsthm}%
\usepackage{mathrsfs}%
\usepackage[title]{appendix}%
\usepackage{xcolor}%
\usepackage{textcomp}%
\usepackage{manyfoot}%
\usepackage{booktabs}%
\usepackage{algorithm}%
\usepackage{algorithmicx}%
\usepackage{algpseudocode}%
\usepackage{listings}%
\usepackage{siunitx}
\usepackage{float}
\usepackage{caption}
\usepackage{subcaption}
\usepackage{hyperref}
\usepackage{url}

\usepackage{times}  
\usepackage{helvet}  
\usepackage{courier}  
\usepackage{comment}
\usepackage{makecell}
\usepackage{mathtools}
\usepackage{scalerel}
\usepackage{multirow,multicol}
\usepackage{xparse,wrapfig}

\usepackage{newfloat}

\theoremstyle{thmstyleone}%
\theoremstyle{thmstyletwo}%

\theoremstyle{thmstylethree}%

\begin{document}

\title[]{Voltage-Controlled Magnetic Tunnel Junction based ADC-less Global Shutter Processing-in-Pixel for Extreme-Edge Intelligence}



\author*[1]{\fnm{Md Abdullah-Al} \sur{Kaiser}}\email{mkaiser8@wisc.edu}
\equalcont{These authors contributed equally to this work.}
\author[2]{\fnm{Gourav} \sur{Datta}}
\equalcont{These authors contributed equally to this work.}
\author[3]{\fnm{Jordan} \sur{Athas}} 
\author[3]{\fnm{Christian} \sur{Duffee}} 
\author[4]{\fnm{Ajey} \sur{P. Jacob}}
\author[3]{\fnm{Pedram} \sur{Khalili Amiri}} 
\author[2]{\fnm{Peter} \sur{A. Beerel}}
\author[1]{\fnm{Akhilesh} \sur{R. Jaiswal}}

\affil[1]{\orgdiv{Electrical and Computer Engineering}, \orgname{University of Wisconsin--Madison}, \orgaddress{\city{Madison}, \state{Wisconsin}, \country{USA}}}

\affil[2]{\orgdiv{Electrical and Computer Engineering}, \orgname{University of Southern California}, \orgaddress{\city{Los Angeles}, \state{California}, \country{USA}}}

\affil[3]{\orgdiv{Electrical and Computer Engineering}, \orgname{Northwestern University}, \orgaddress{\city{Evanston}, \state{Illinois}, \country{USA}}}

\affil[4]{\orgdiv{Information Sciences Institute}, \orgname{University of Southern California}, \orgaddress{\city{Los Angeles}, \state{California}, \country{USA}}}

\abstract
{The vast amount of data generated by camera sensors has prompted the exploration of energy-efficient processing solutions for deploying computer vision tasks on edge devices. Among the various approaches studied, processing-in-pixel integrates massively parallel analog computational capabilities at the extreme-edge \textit{i.e.,} within the pixel array and exhibits enhanced energy and bandwidth efficiency by generating the output activations of the first neural network layer rather than the raw sensory data. In this article, we propose an energy and bandwidth efficient ADC-less processing-in-pixel architecture. This architecture implements an optimized binary activation neural network trained using Hoyer regularizer for high accuracy on complex vision tasks. In addition, we also introduce a global shutter burst memory read scheme utilizing fast and disturb-free read operation leveraging innovative use of nanoscale voltage-controlled magnetic tunnel junctions (VC-MTJs). Moreover, we develop an algorithmic framework incorporating device and circuit constraints (characteristic device switching behavior and circuit non-linearity) based on state-of-the-art fabricated VC-MTJ characteristics and extensive circuit simulations using commercial GlobalFoundries 22nm FDX technology. Finally, we evaluate the proposed system's performance on two complex datasets - CIFAR10 and ImageNet, showing improvements in front-end and communication energy efficiency by $8.2{\times}$ and $8.5{\times}$ respectively and reduction in bandwidth by $6{\times}$ compared to traditional computer vision systems, without any significant drop in the test accuracy.}


\maketitle

\section{Introduction}\label{intro} 
Today's widespread computer vision applications have spurred the development of energy-efficient processing and computing solutions \cite{auto_driving, surveillance, obj_track, anomaly_detect, vid_obg_detect}. Traditionally, CMOS image sensors (CIS) convert the visual scene captured by a 2D pixel array into multi-bit digital data utilizing energy and delay intensive analog-to-digital converters (ADC) \cite{cis1, cis2}. Hence, in conventional computer vision (CV) systems, a significant amount of data per frame (determined by the number of pixels inside the camera sensors multiplied by the ADC bit precision) needs to be transmitted off-chip for further backend processing and artificial intelligence (AI) applications. This physically segregated approach of sensing in the camera and processing elsewhere results in energy, throughput, and bandwidth bottlenecks. Hence, considerable interest has grown within the research community for developing novel energy-efficient hardware solutions for resource-constrained edge devices deployed for computer vision applications. 

To address these bottlenecks, researchers are exploring various solutions to enhance energy and bandwidth efficiency by bringing computational tasks closer to the sensor array \cite{in_sensor_mv, near_in_sensor_survey}. These approaches can be categorized into three classes: (a) near-sensor processing, (b) in-sensor processing, and (c) in-pixel processing. 
In near-sensor processing, digital signal processors (DSPs) or machine learning (ML) accelerators are placed on a different chip/die adjacent to the sensor, either on the same PCB or in a 3D chip stack \cite{near_sensor_ARVR, near_sensor_3D_sony}. Consequently, near-sensor processing enhances energy efficiency by reducing data transfer costs between the sensor and the cloud; however, data flow between the sensor and the near-sensor processor still encounters bandwidth and energy bottlenecks.
In contrast, in-sensor processing utilizes an analog or digital signal processor at the periphery of the sensor chip, resulting in a significant reduction in off-chip data transfer costs \cite{mixed_mode_ivs, analog_bnn_swcap, sleepspotter}. This approach still requires transferring raw analog sensor data through column-parallel bitlines to the peripheral processing networks. Hence, the in-sensor processing approach still encounters data transfer bottlenecks between the sensor and peripheral processor.
On the other hand, the in-pixel processing approach integrates computation capabilities inside the sensor/pixel array, minimizing data transfer overhead significantly \cite{senputing, pwm_idac_weight, aps_p2m, aps_p2m_detrack}. The in-pixel processing elements perform computation on the raw analog sensor data and generate the feature activation output per kernel of the first layer of the neural network (NN). Thus, this approach results in substantial improvements in energy efficiency and throughput, as well as significant reductions in bandwidth.

In short, in-pixel processing yields improved energy and bandwidth efficiency compared to traditional and other computer vision edge systems such as in-sensor and near-sensor processing. This improvement is achieved due to the reduction of the number of ADC operation in the in-pixel solution due to its kernel-level readout approach, requiring only one conversion step for each kernel, say with a size of 3x3x3, while other CV approaches may necessitate up to 27 ADC conversions for the same kernel size. Additionally, current NN algorithms, such as quantization-aware training, contributes to this efficiency by achieving high accuracy with fewer ADC bits than those needed for raw sensory data. However, the use of the multi-bit representation for output activations generated by the multi-bit ADCs, that are both energy and delay hungry, in the in-pixel approach limits the achievable energy consumption and bandwidth improvement \cite{aps_p2m,aps_p2m1}. Moreover, complex machine vision tasks require multiple channels in the first NN layer \cite{vgg,resnet} which can exacerbate the rolling shutter effect, where rows of pixels are exposed sequentially rather than simultaneously \cite{dai2016rolling,fan2021sunet}. Additionally, rolling shutter effect needs to be corrected in the NN pipeline, particularly for fine-grained machine vision tasks, otherwise the accuracy drop is significant \cite{hu2023lets}. Due to the multi-cycle sequential rolling shutter operation per channel, the in-pixel computing approach can intensify the adverse effects of rolling shutter and lead to motion blur, impacting image quality more severely than in conventional systems.

In this work, we propose for the first time, an ADC-less binary activation neural network (NN) architecture to further optimize the energy and bandwidth efficiency of the in-pixel approach without any significant drop in end application accuracy. At first, we multiply the analog input (incident light intensity per pixel of a visual scene) with the weight (coefficient of the algorithmic kernel) and compute the binary feature activation spike. Consequently, we reduce the data bandwidth by utilizing single-bit activation spikes, implemented through innovative use of VC-MTJs, instead of multi-bit data, thereby significantly improving the bandwidth efficiency. In addition, we achieve enhanced energy efficiency by employing a passive analog subtractor and a threshold operation instead of the multi-bit energy-intensive ADC used in prior works. Furthermore, the output activations of the binary activation NN are sparse; hence, the architecture can also reduce the overall activation switching energy. 

In addition, to alleviate the motion blur effects caused by the rolling shutter mechanism, we propose implementing a global shutter operation with burst-read scheme for output feature activations utilizing voltage-controlled magnetic tunnel junctions (VC-MTJs). Compared to other non-volatile memory (NVM) devices, including memristors, resistive random access memory (RRAM), and phase change memories (PCM) \cite{rram_endurance, pcm_endurance, memristor_endurance}, MTJ devices demonstrate enhanced endurance and hence are amenable to the proposed processing-in-pixel scheme that requires multiple read-write cycles per exposure \cite{mram_endurance}. Additionally, our fabricated VC-MTJ device showcases near-deterministic switching within sub-nanosecond time frames (compared to typical microsecond pixel integration time) that significantly reduces the latency overhead associated with the use multiple VC-MTJ devices, described in detail in the following sections. Furthermore, leveraging the voltage-controlled magnetic anisotropy (VCMA), our VC-MTJ can switch with less than 1V applied bias, which aligns well with advanced technology nodes (specifically, we utilize commercial GlobalFoundries 22 nm FD-SOI technology for our hardware verification). Moreover, due to the VCMA effect, by reversing the polarity of read voltage compared to the write voltage, disturbance-free read operations can be achieved. This enables the implementation of our proposed high-speed burst-read scheme. Note, the switching threshold of the VC-MTJ is predetermined based on the device characteristics and may not align with the threshold required by the NN algorithm. Moreover, the threshold can vary based on the network architecture (e.g. variants of VGG and ResNet) and the dataset trained (e.g. CIFAR10 and ImageNet). To accommodate these algorithmic needs, we introduce a novel algorithm-hardware tunable mapping scheme to match the algorithm equivalent hardware threshold with the VC-MTJ's switching requirement by repurposing the analog subtractor block. Given its high endurance, faster write speeds, disturbance-free reads, and non-volatile nature, along with the novel tunable threshold matching techniques, VC-MTJ emerges as a promising candidate for our processing-in-pixel binary activation NN system exhibiting significant energy and latency improvement without any significant drop in the test accuracy for state-of-the-art complex CV datasets. Importantly, the global shutter operation inevitably increases the frames per second (FPS) of our processing-in-pixel architecture compared to existing in-sensor and baseline CV systems, thereby enabling resource-constrained real-time applications.

The key contributions of our work are as follows:

\begin{enumerate}
\item We propose a novel \textit{high-speed global shutter processing-in-pixel} scheme by exploiting high endurance, high write speed, disturb-free read and non-volatility of state-of-the-art nanoscale VC-MTJs enabling massively parallel analog computing for extreme-edge intelligence.
\item Moreover, we present \textit{ADC-less in-pixel architecture} by employing a passive analog subtractor circuit to compute the feature spike of the first binary activation NN layers using thresholding and the storage characteristics of VC-MTJs.
\item Furthermore, we introduce a \textit{novel tunable mapping scheme} to align the algorithm equivalent hardware threshold with the VC-MTJ's switching behavior by repurposing the analog subtractor block and utilize multiple VC-MTJ neurons to enable high-confidence output activation computation that is critical to meet accuracy demands. 
\item Finally, we develop a \textit{device-circuit-algorithm co-design framework} considering device and hardware constraints, and evaluate the system performance utilizing CIFAR10 and ImageNet datasets. Our algorithmic framework utilizes measured device characteristics for state-of-the-art fabricated VC-MTJs and HSpice simulations on commercial GlobalFoundries 22nm FDX technology. 
\end{enumerate}

\section{Methods}\label{methods}

\subsection{VC-MTJ as enabler for ADC-less In-pixel Global Shutter Operation:}\label{device_details}  

\begin{figure}[!t]
\centering
\subfloat[]{\includegraphics[width=0.3\linewidth]{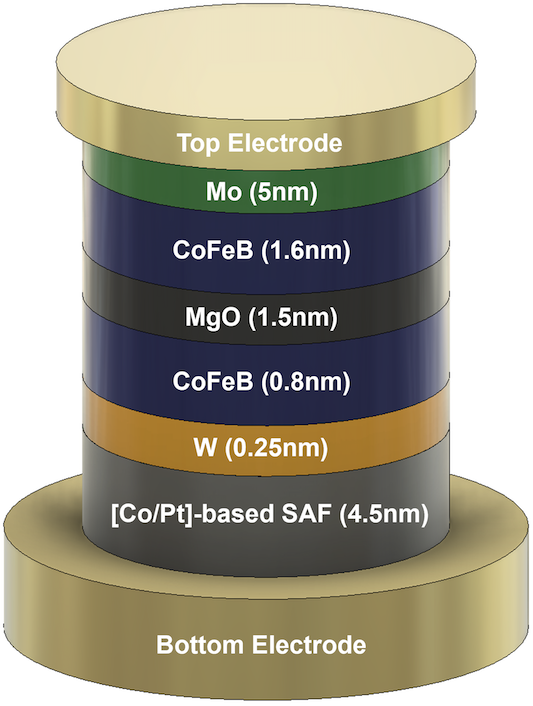}} \hspace{0.1\textwidth}
\subfloat[]{\includegraphics[width=0.5\linewidth]{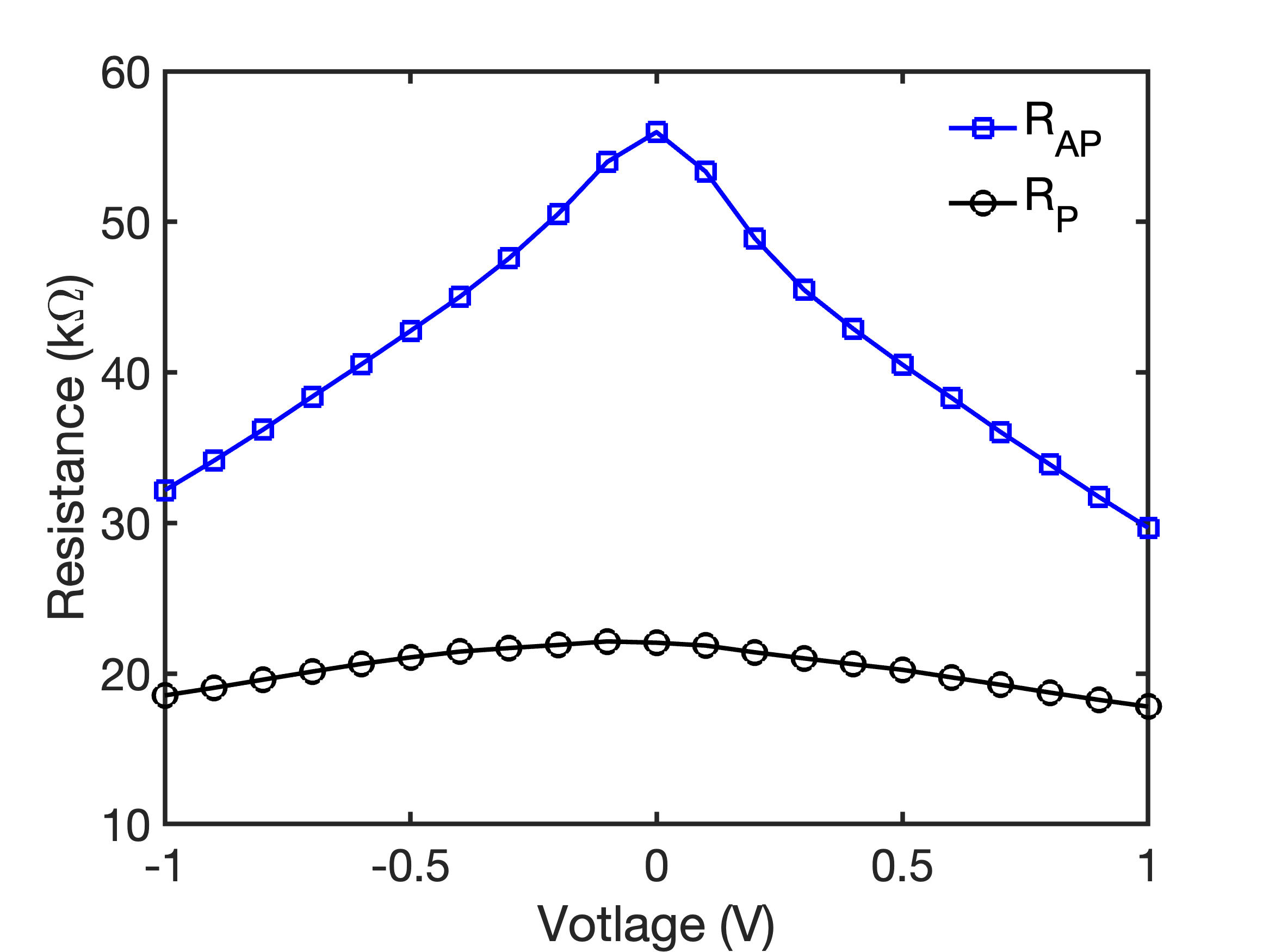}}
\caption{(a) The stack of the MTJs used in this work. Each MTJ is patterned as a circular pillar with a diameter of approximately 70 nm. (b) Dependence of the parallel (\si{R_P}) and antiparallel (\si{R_{AP}}) state resistances on voltage applied across the device. A TMR in excess of 150\% is found at near-zero voltages.}
\label{stack}
\end{figure}

A model of the VC-MTJ stack used in this work is depicted in Fig. \ref{stack}(a). The device structure comprises a bottom electrode and pinning layer, realized from a Co/Pt-based synthetic antiferromagnetic (SAF) layer/\si{Co_{20}Fe_{60}B_{20}} (0.8)/MgO (\si{\sim}1.5)/\si{Co_{17.5}Fe_{52.5}B_{30}}(1.6)/Mo (5)/top electrode, where the numbers in parentheses denote thickness in nanometers. The top CoFeB layer thickness is designed to ensure perpendicular magnetic anisotropy (PMA) of its magnetization orientation  \cite{Perp_CoFeB_MgO}. A Mo capping layer is used to enhance thermal annealing stability by preventing diffusion into the top CoFeB layer. The B concentration in the bottom CoFeB layer was chosen to be 20\%, which has previously shown to promote the formation of (001) MgO, leading to an increased tunneling magneto-resistance (TMR) \cite{ikeda2012boron}. The top CoFeB layer is richer in B, which promotes enhanced VCMA after high-temperature annealing, as reported in a previous work \cite{vcma_switching, VC_MTJ_Persp}. The devices selected for this application were patterned into 70 nm diameter pillars, as previous studies have exhibited sub-volt VCMA switching at this size \cite{vcma_switching}.

For the following measurements, we define positive voltage as being applied from the top device electrode to the bottom electrode. Fig. \ref{stack}(b) illustrates the device resistance as a function of DC voltage applied across the device, ranging from -1V to +1V. The low and high resistance levels correspond to the magnetization orientations of the MTJ's top CoFeB (free layer) being either parallel (P) or anti-parallel (AP) to that of the bottom CoFeB (reference layer). The bottom CoFeB layer magnetization orientation is fixed by the pinning stack, allowing only the magnetization direction of the free layer to be switched. When measuring \si{R_{AP}} and \si{R_P}, a bias out-of-plane magnetic field was applied to force the magnetization direction of the free layer into the desired P or AP state. The TMR ratio, defined as \si{(R_{AP} - R_P)/R_P}, shows a value \si{>}150\% at near-zero readout voltages (1 mV). Larger applied voltages, positive or negative, lead to a decrease in \si{R_{AP}} as typically observed in MgO-based tunnel junctions. The large TMR value of \si{>}150\% is crucial to enable the burst read scheme facilitating the global shutter operation, as discussed latter.

\begin{figure}[!t]
\centering
\subfloat[]{\includegraphics[width=0.5\linewidth]{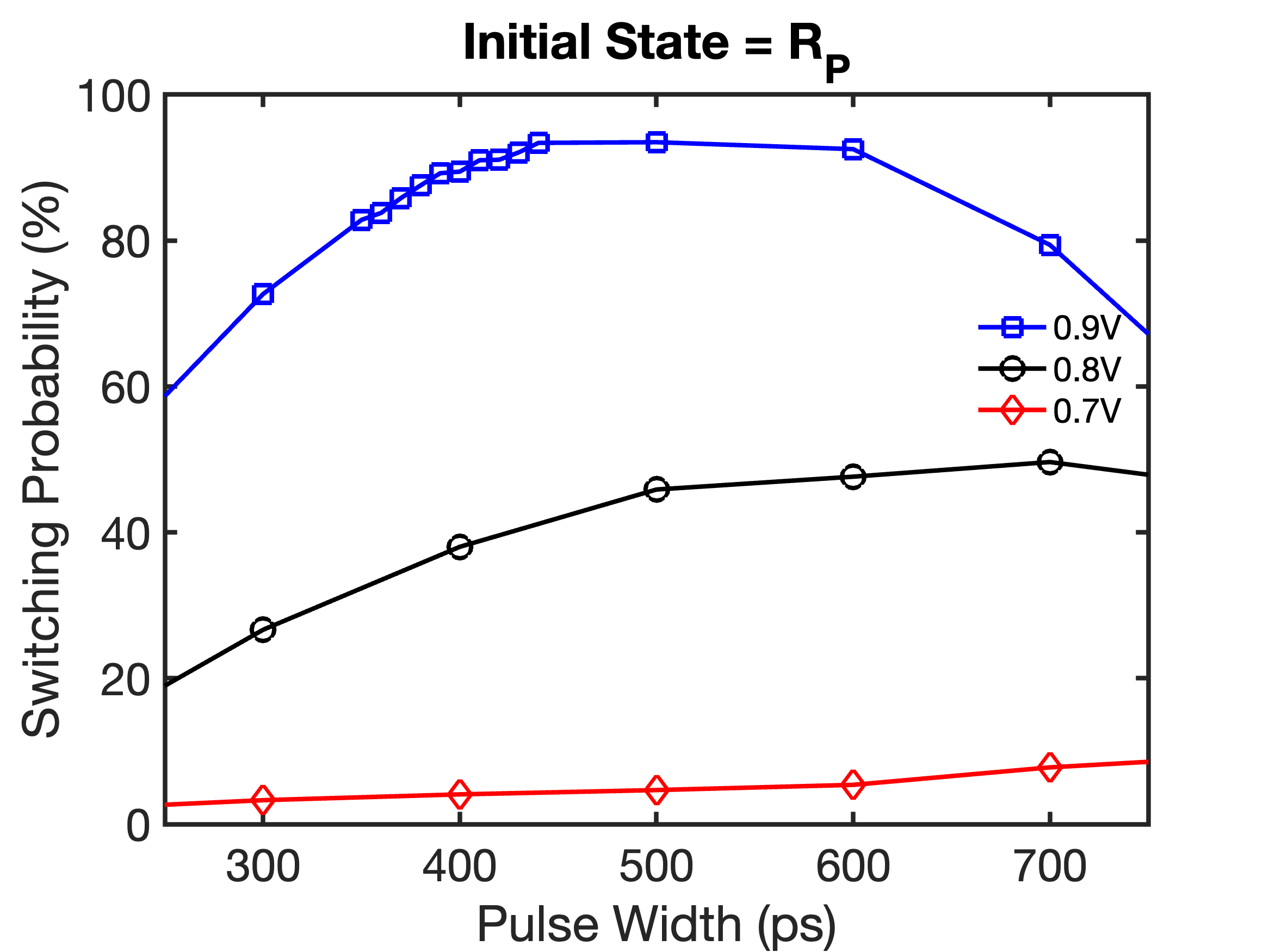}} 
\subfloat[]{\includegraphics[width=0.5\linewidth]{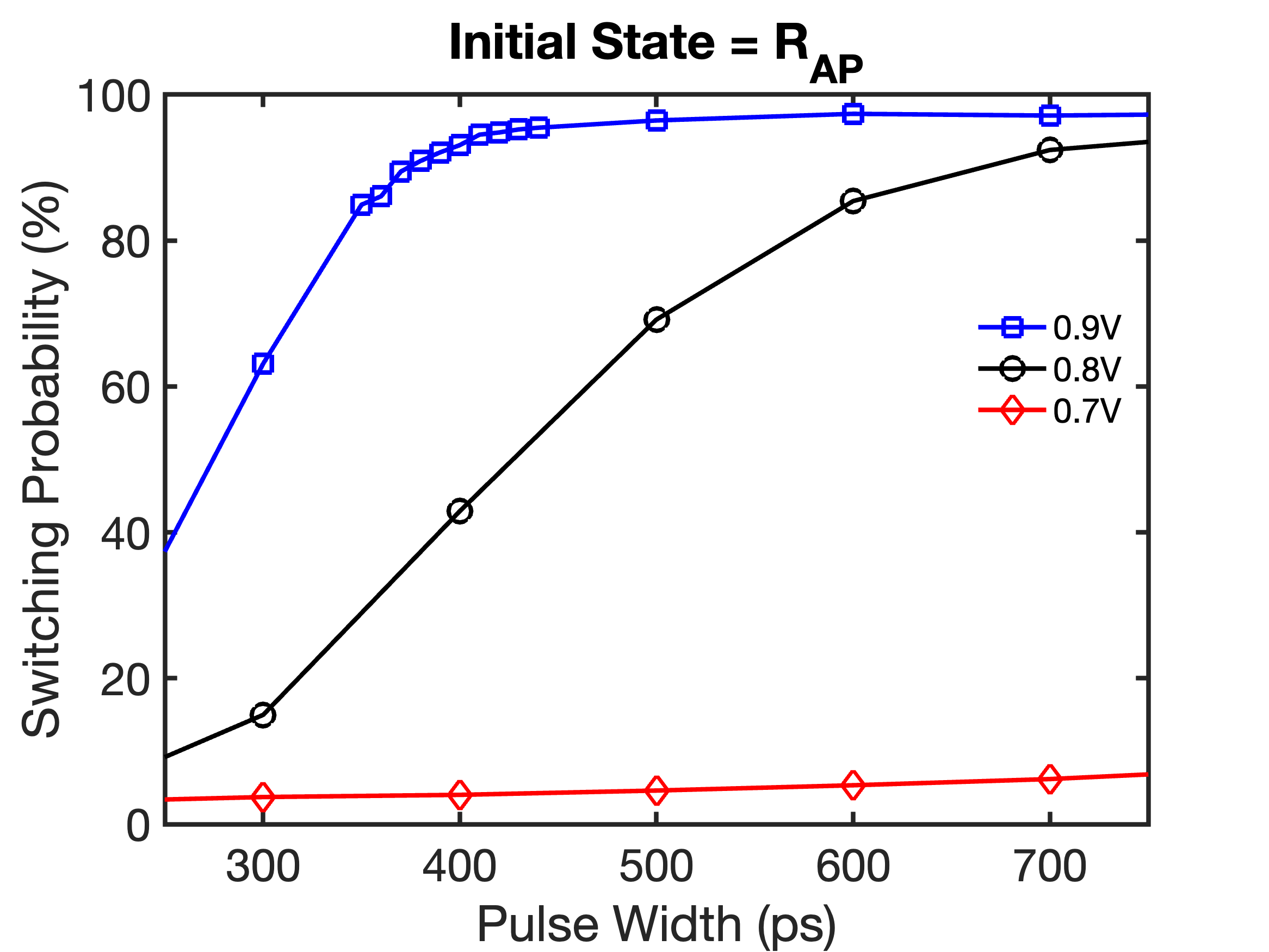}}
\caption{Switching probability is shown as a function of pulse widths at various applied voltages (ranging from 0.7V to 0.9V), with the initial resistance state being (a) parallel (\si{R_P}) and anti-parallel (\si{R_{AP}}). For VC-MTJs to function effectively as binary thresholding neurons, they are expected to switch with high probability at higher voltages. Therefore, the (\si{R_{AP}}) state is used as the reset state for the binary neurons, as it demonstrates a higher switching probability across a wide range of pulse widths.}
\label{switch}
\end{figure}

We leverage the VCMA effect to achieve near-deterministic switching of the MTJ's free layer in response to short voltage pulses \cite{VCMA_spintronics, EFC_RAM, low_energy_nano}. Due to the VCMA effect, the voltage reduces the device's energy barrier between the \si{R_P} and \si{R_{AP}} sufficiently to induce precession around an in-plane bias magnetic field (which can, in principle, be built into the device using a modified fixed layer). Upon removal of the VCMA voltage pulse, PMA is restored, and the free layer's magnetization solidifies along the closest perpendicular direction. This phenomenon can be utilized to switch the free layer magnetization by adjusting the duration of the applied voltage pulse to be multiples of half the period of a full precession. Fig. \ref{switch}(a) and Fig. \ref{switch}(b) depict the switching probability of the MTJ as a function of pulse duration with the initial state being  parallel and antiparallel, respectively. It can be observed that near-deterministic switching can be achieved in both cases with sub-nanosecond voltage pulses, showcasing the remarkably high write speed of these VC-MTJs. The same measurement was conducted with decreasing voltage amplitudes, revealing a decrease in switching probability as voltage magnitude decreases, eventually dropping to near-zero switching within a few hundred mV. Note, in order to use VC-MTJs as binary thresholding neurons they are expected to switch at higher voltages and not switch at lower voltages. As shown in Fig. \ref{switch}(b), the switching probability exceeds 90\% when the applied voltage is above 0.8 V at pulse width of 700 ps.

Due to the substantial energy barriers between states, the MTJ acts as a non-volatile memory, unless its VCMA voltage threshold is exceeded, forgoing the need for an energy-intensive refresh cycle \cite{li2018voltage}. Moreover, using voltage polarity such that the PMA (energy barrier) increases due to the VCMA effect allows disturb-free read operation, which is leveraged to enable high-speed burst read operation for the pixel array. Coupled with their significant TMR ratio, minimal required area, rapid switching speed, and disturb-free read operation, VC-MTJs emerge as an ideal choice for the proposed global shutter, ADC-less processing-in-pixel scheme.

\subsection{In-pixel Computing Circuit featuring Device Redundancy and Switching Threshold Matching Scheme}\label{inpix_ckt} 

\begin{figure}[!t]
\centering
\subfloat{\includegraphics[width=0.5\linewidth]{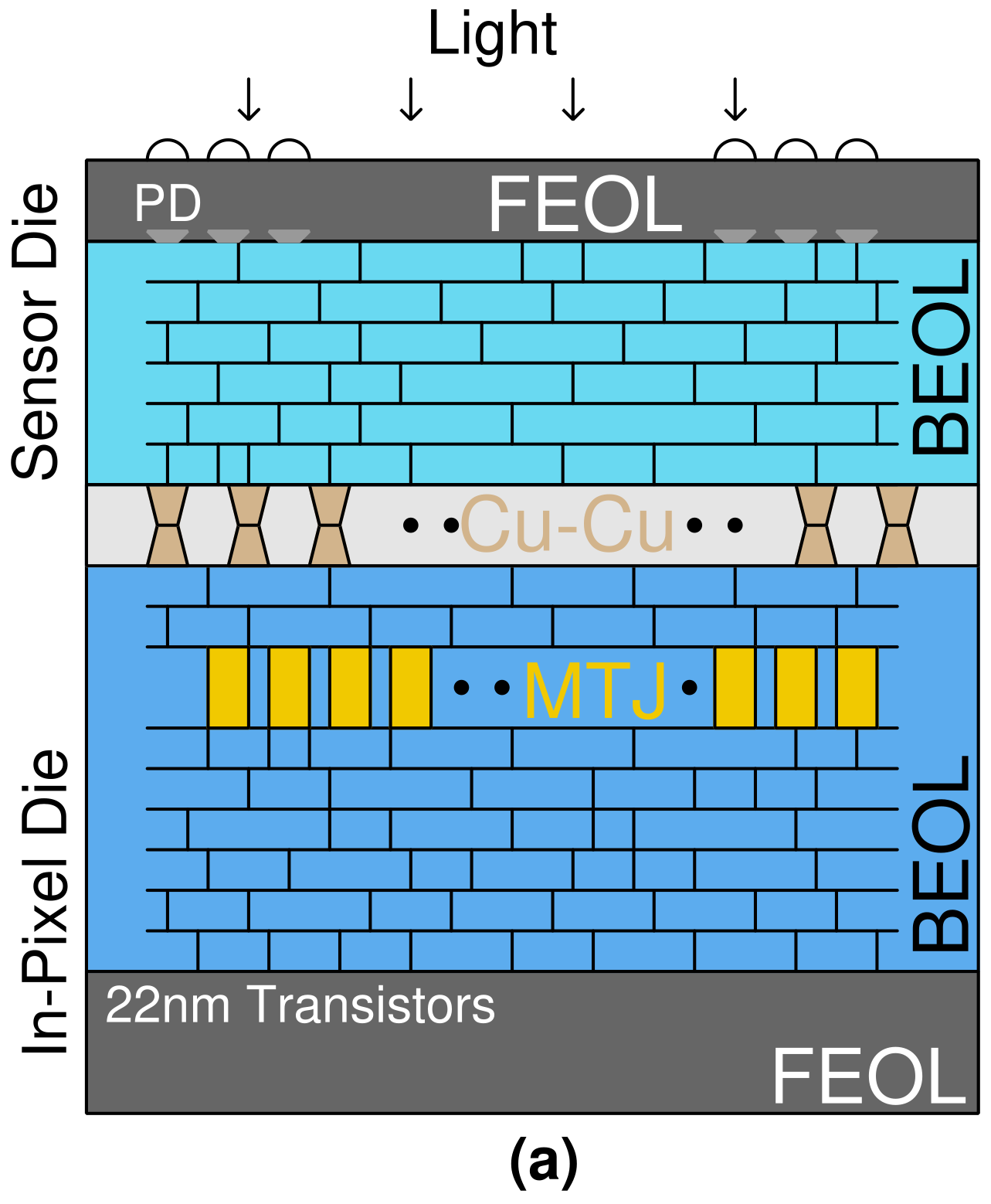}}  \\
\subfloat{\includegraphics[width=0.9\linewidth]{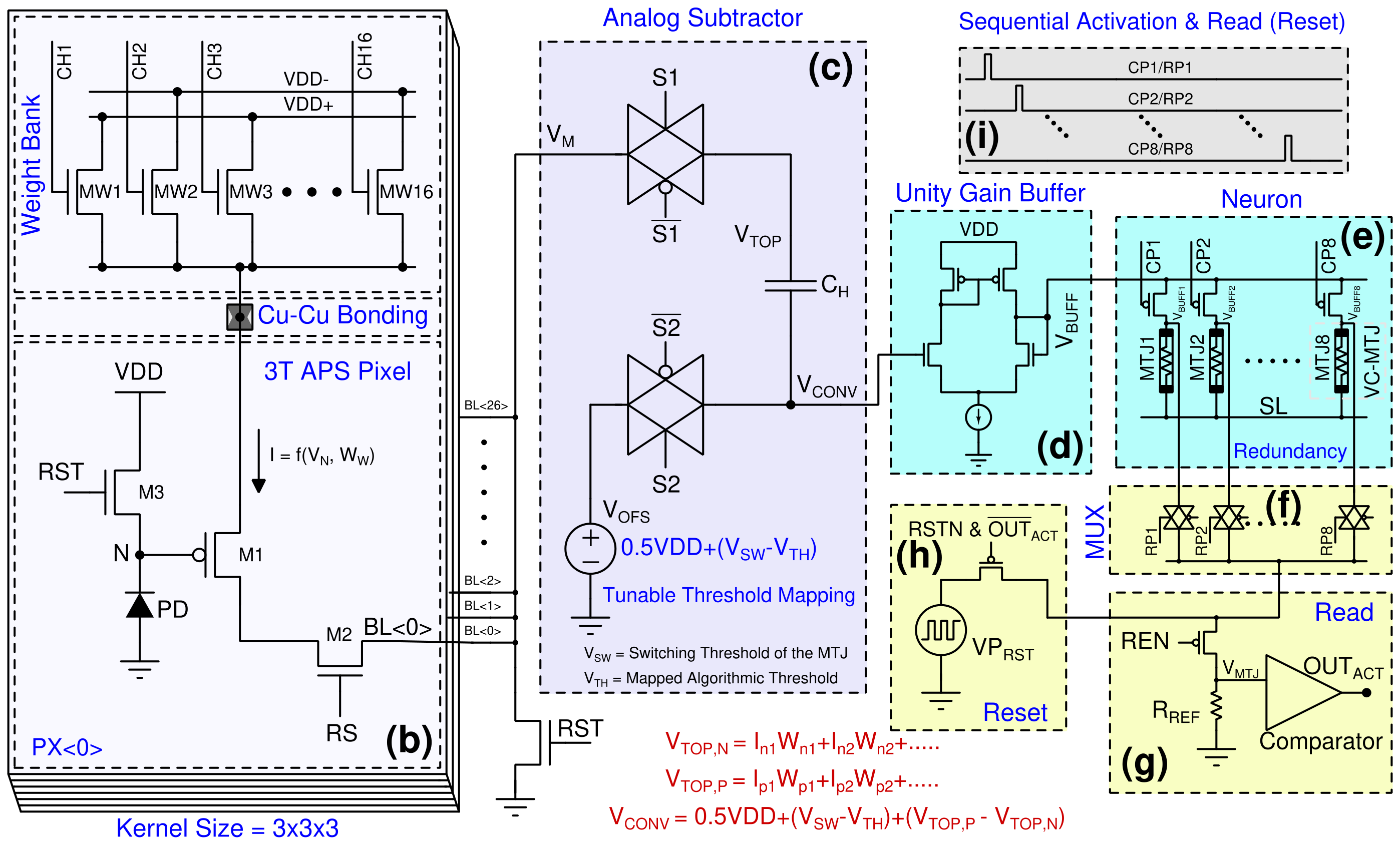}}
\caption{(a) Representative illustration of heterogeneously 3D integration of our proposed processing-in-pixel solution utilizing Cu-Cu hybrid bonding, where the top die comprises the pixel sensor array, and the bottom die consists of the processing blocks. Circuit implementation of (b) weight-augmented pixel circuit, (c) analog subtractor to compute the final convolution output, (d) unity-gain buffer to drive to VC-MTJ neurons, (e) multiple VC-MTJ neurons, (f) MUX for selective read and reset, (g) comparator-based read circuits, (h) reset pulse generator, (i) control pulses for burst mode write and read operation.}
\label{circuit}
\end{figure}

In this subsection, we will discuss all computational aspects of our proposed processing-in-pixel solution. Figure \ref{circuit} illustrates the hardware implementation of our proposed ADC-less global shutter processing-in-pixel approach. The operations of our proposed approach can be divided into three phases. These are:-
\begin{enumerate}
    \item Convolution Phase: In this phase, after a pre-define integration time, weight-augmented pixel circuits generate multiply-and-accumulate (MAC) results for positive and negative weights sequentially over two integration cycles. Each integration cycle is preceded by a photodiode reset operation using RST signal. An analog subtractor estimates the final analog convolution output.
    \item Binary Activation Phase: An unity-gain buffer sequentially drives the multiple VC-MTJs acting as neurons at the analog convolution output (V$_{CONV}$) of the previous phase and generates the binary activation (switching resistance state from anti-parallel to parallel) if applied bias exceeds the switching threshold.
    \item Burst Read and Reset Phase: Finally, the multiple VC-MTJs are read in a burst manner using a comparator, followed by a reset operation (anti-parallel state of the VC-MTJs) if needed. 
\end{enumerate}

More details on each step, including their hardware implementations, is explained below.

\subsubsection{Weight-augmented Pixel Circuit:} 
Our weight-augmented pixel circuit comprises a 3T pixel configuration and additional transistors representing the weights of the neural network, as shown in Figure \ref{circuit}(b). The algorithmic weight value is implemented by modulating the geometry of the weight transistors (MW1, MW2, etc.), with larger weight values corresponding to wider transistors. Additionally, positive and negative weights are implemented by connecting the weight transistors to VDD+ (supply for the positive weights) and VDD- (supply for the negative weights), respectively. Moreover, complex vision tasks necessitate multiple NN channels; hence, multiple weight transistors are connected per pixel. Each weight transistor can be activated independently by its gate control signals (CH1, CH2, etc.). One of the key enablers of embedding multiple weight transistors per pixel to support the number of algorithmic channel requirements is the advancement of heterogeneous 3D integration \cite{3D_samsung, 3D_sony}. Our proposed hardware can be heterogeneously 3D integrated utilizing fine-pitch Cu-Cu hybrid bonding \cite{cu2cu_pitch}, where the top die comprises the pixel sensor array, and the bottom die consists of the weight transistors with additional compute circuits (as shown in Fig. \ref{circuit}(a)), aligned vertically with each pixel group per kernel.

The pixel circuit comprises a photodiode that converts incident light into electrical current. Initially, the reset transistor (M3) is activated, resetting the voltage of the internal node (N) to VDD. During the convolution phase, the node N discharges depending on the intensity of the incident light, discharging faster with higher light intensity and slower with lower light intensity. The current produced by transistor M1 is influenced by the gate voltage (node N) and is proportional to the incident optical input. The weight transistors are connected to the source terminal of the input transistor M1; therefore, these weight transistors introduce source degeneration effects. Higher width (lower resistance) in these weight transistors results in higher modulated current values. Thus, the current passing through transistor M1 is influenced by both the gate voltage (input light intensity) and the weight (width or resistance of the weight transistor).

Figure \ref{scatter_plot_sub}(a) illustrates the normalized simulated output voltage corresponding to a 3x3x3 kernel size plotted against the normalized input intensity multiplied by the weight. The voltage range of the hardware is linearly mapped to the algorithmic normalized range of [-3,3]. The simulations were conducted using the GlobalFoundries 22 nm FDX technology node across various combinations of input intensity ranges and weights. It is evident from the figure that the simulated output (depicted by blue circles) closely tracks the ideal convolution (represented by the linear black line), albeit with some non-linear effects stemming from the transistors. This circuit characteristic is incorporated in the algorithm by substituting the ideal convolution function with a hardware-aware custom convolution function to accurately capture these effects (discussed in  Subsection \ref{codesign_details}).

\subsubsection{Analog Convolution Operation and Threshold Matching Scheme:}
We perform the subtraction of the convolution output for negative and positive weights in two distinct phases to estimate the final convolution output using the passive analog subtractor circuit shown in Fig \ref{circuit}(c). During phase-1, we activate the negative weight transistors within the kernel by enabling corresponding gate control signals, resulting in the generation of the multiplication terms (input light intensity\si{\times}weight) inside the weight-augmented pixels, which are then combined through the shared bitlines (node \si{V_M}). In this phase, both S1 and S2 switches (shown in Fig \ref{circuit}(c)) are activated, causing the multiply-and-accumulate (MAC) output of the negative weights per kernel to be stored (\si{V_{TOP}} follows the \si{V_M}) on the top plate of the storage capacitor (\si{C_H}), while the bottom plate of the capacitor is charged with the DC voltage. Subsequently, in phase-2, we only activate the positive weight transistors and execute the MAC operation for positive weights only. In this phase, only the S1 switch is activated, causing the top plate of the capacitor (\si{V_{TOP}}) to track the voltage on \si{V_M}. The change in voltage (difference between the analog voltage for negative and positive weights) on the top plate of capacitor is coupled to the bottom plate of the capacitor. As a result, the bottom plate generates the subtracted output from the MAC results of negative to positive weights with a DC offset. Thus, the final analog convolution output (\si{V_{CONV}}) is computed using the passive analog subtractor circuit.

\begin{figure}[!t]
\centering
\subfloat[]{\includegraphics[width=0.75\linewidth]{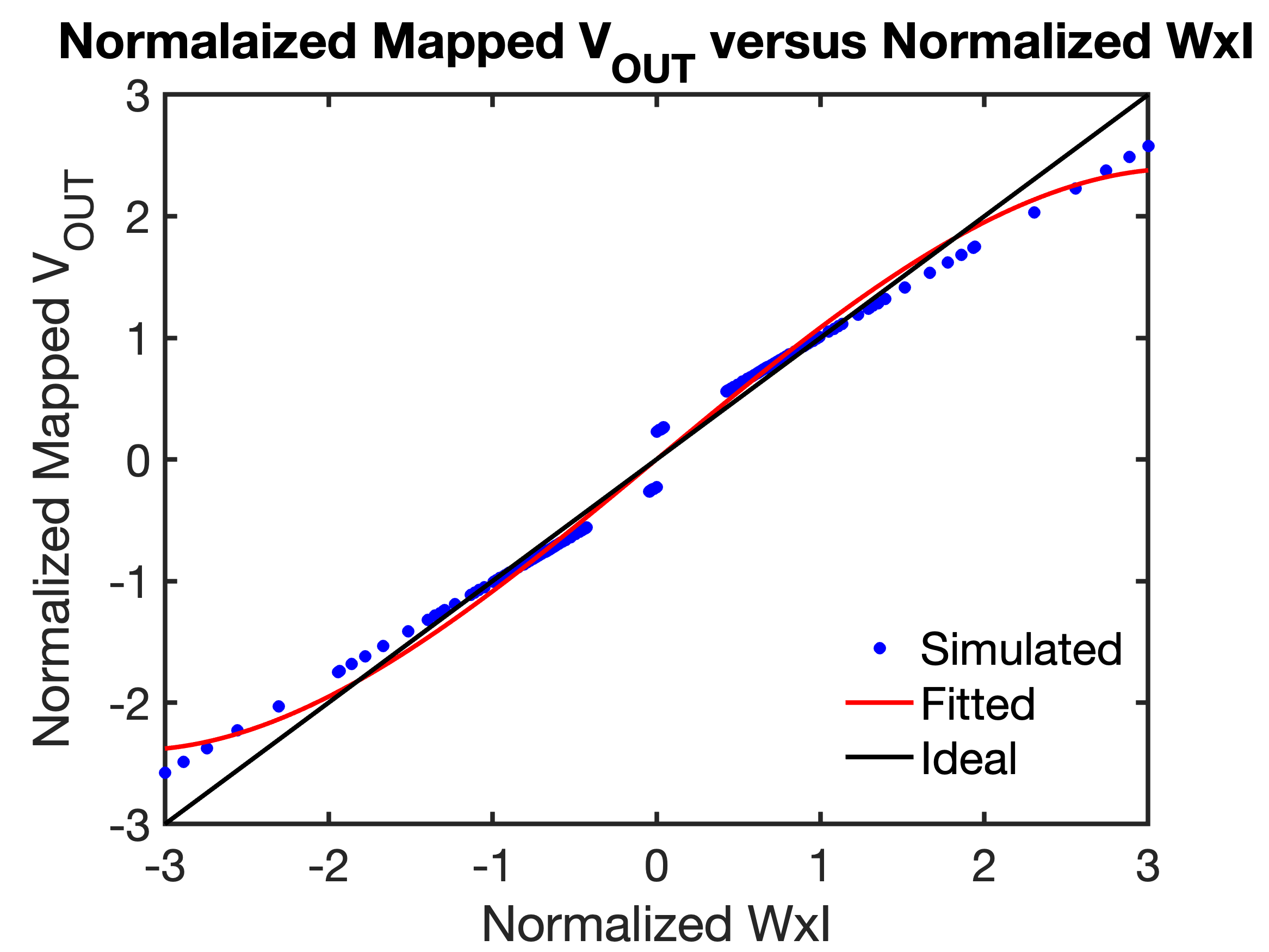}} \\
\subfloat[]{\includegraphics[width=0.75\linewidth]{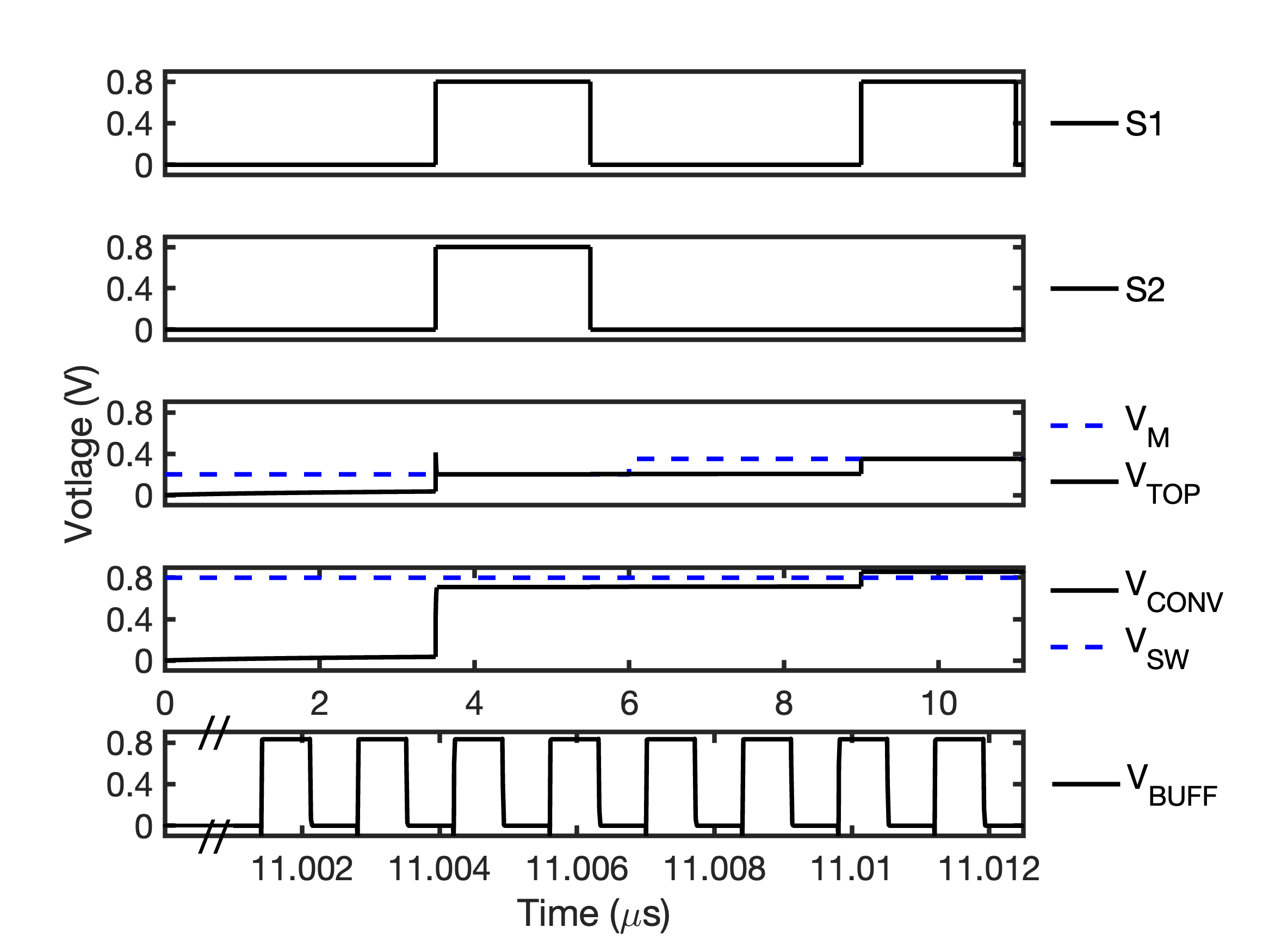}}
\caption{(a) A scatter plot comparing the normalized weight-augmented pixel output voltages to the ideal normalized multiplication values of various weights and input activations (Normalized W \si{\times} I). (b) Computation of final convolution output and burst-write driving pulses for binary activations of VC-MTJ neurons.}
\label{scatter_plot_sub}
\end{figure}

Note, we employ a binary activation neural network (as described in the Subsection \ref{algo_details}), which involves comparing the analog convolution output per kernel with a predefined threshold value to generate output the binary activation. If the output exceeds the threshold, the activation is set to 1; otherwise, it remains at 0. Our hardware utilizes VC-MTJ both as neurons and for enabling global shutter computing operations across the 2D array due to its non-volatile nature. As shown in Fig. \ref{switch}(b), near-deterministic switching is achieved when the applied 700 ps wide voltage pulse across the VC-MTJ exceeds 0.8V with the error percentage below $\sim$10\% with the initial state of the VC-MTJ being antiparallel. Conversely, as shown in Fig. \ref{switch}(b), when the applied 700 ps wide voltage pulse across the VC-MTJ is below 0.7V the error percentage is again below $\sim$10\%. Thus, we aim to align the near-deterministic switching of device above 0.8V (a predetermined device-dependent threshold) with the algorithmically required threshold by introducing novel threshold matching technique by repurposing the analog subtractor circuit in our hardware. Typically, the source DC voltage (\si{V_{OFS}}) is set to 0.5\si{VDD} to maintain an equal dynamic range for the subtraction operation. However, to incorporate the switching threshold of the VC-MTJ, we adjust the offset voltage to 0.5\si{VDD+(V_{SW}-V_{TH})}, where \si{V_{SW}} represents the switching voltage of the VC-MTJ and \si{V_{TH}} is the hardware-mapped algorithmic threshold voltage. Typically, the \si{V_{SW}} is larger than the \si{V_{TH}}, hence, the DC offset is skewed toward VDD. Furthermore, this global DC offset can be externally controlled, enabling the adjustment of the threshold requirement by the algorithm. In addition, the skewed offset will not impact the final activation generation as long as it exceeds the switching threshold voltage of the VC-MTJ, even if the analog convolution output saturates due to the skewed offset. This is because once the analog output surpasses the threshold, a binary activation is generated irrespective of the amount by which the analog output exceeds the threshold. Thus, by repurposing the analog subtractor circuit, we implement novel technique for tunable threshold mapping, independent of the predetermined threshold voltage dictated by VC-MTJ's switching behavior.

Figure \ref{scatter_plot_sub}(b) illustrates the transient behavior of calculating the analog convolution and burst-write operation of the multiple VC-MTJs. During phase-1, the MAC results of the negative weights are accumulated on the top plate of the capacitor when S1 is activated (\si{V_{TOP}} follows \si{V_M}). In this phase, S2 is also active, causing the bottom plate of the capacitor (\si{V_{CONV}}) to track the DC offset (\si{V_{OFS}}). The DC voltage is adjusted to ensure that the hardware equivalent algorithmic threshold exceeds the VC-MTJs switching threshold, thereby ensuring high-confidence switching. In phase-2, only S1 is activated while S2 remains off. Consequently, the floating bottom plate tracks the voltage difference in the top plate, causing the VC-MTJs to switch from their reset state (anti-parallel) to parallel state sequentially if the convolution output (\si{V_{CONV}}) exceeds the switching threshold of the VC-MTJs, as depicted in Figure \ref{scatter_plot_sub}(b). The buffered voltage (\si{V_{BUFF}}) is used to drive the multiple VC-MTJs sequentially following the control pulses (CP1, CP2, etc.) as shown in shown in the Fig. \ref{circuit}(d) and (i). Note, the unity gain buffer it power gated during the convolution and read operation and activated only during the burst-write phase. 

\subsubsection{VC-MTJ Neuron Activation}
Our fabricated VC-MTJ demonstrates a near-deterministic switching (error percentage \si{<10}\%) for an applied voltage of 0.8V shown in the Fig. \ref{switch}(b)). The relationship between algorithmic accuracy and the tolerable error percentage in device switching behavior is discussed in subsection \ref{codesign_details}. Analysis reveals that achieving high accuracy necessitates a high-confidence switching probability (low error percentage $<$2\%). Thus, we employ multi-VC-MTJs as our output binary activation compute units (shown in the Fig. \ref{circuit}(e)), wherein multiple VC-MTJs are switched using the accumulated analog convolution voltage and subsequently a majority function decides the final binary output activation.  The final convolution output from the analog subtractor (\si{V_{CONV}}) is applied to the top electrode of each VC-MTJ followed by a unity gain buffer. We execute sequential write operations on each VC-MTJ independently. During the thresholding phase, the buffered analog convolution output is supplied to the the multi-VC-MTJs (8 VC-MTJs per kernel have been used in this work) by sequentially activating the control signals (CP1, CP2, etc.) with appropriate pulses (700 ps pulse to switch from anti-parallel to parallel state) as depicted in Fig. \ref{circuit}(i). Advantageously, our experimental measurements exhibit fast sub-ns switching (shown in Fig. \ref{switch}(b)) of VC-MTJ, hence, the sequential VC-MTJ writing does not incur significant latency overhead as pixel integration time is in \si{\mu s} range. Further, the back-end-of-line compatible nanoscale size of our VC-MTJ (70nm in diameter) is much smaller compared to typical pixel sizes having pixel pith in micro-meter scale.  The bottom electrodes of the MTJs are interconnected and connected to the source line (SL). This voltage can also be externally regulated by which we can adjust the voltage drop due to loading effect of the VC-MTJs that act as resistive loads. Utilization of the multi-VC-MTJs as the activation generation unit ensure high-confidence switching behavior (with error percentage well below $<$2\% ) which is required by the algorithm for achieving high accuracy. 

\begin{figure}[!t]
\centering
\subfloat[]{\includegraphics[width=0.5\linewidth]{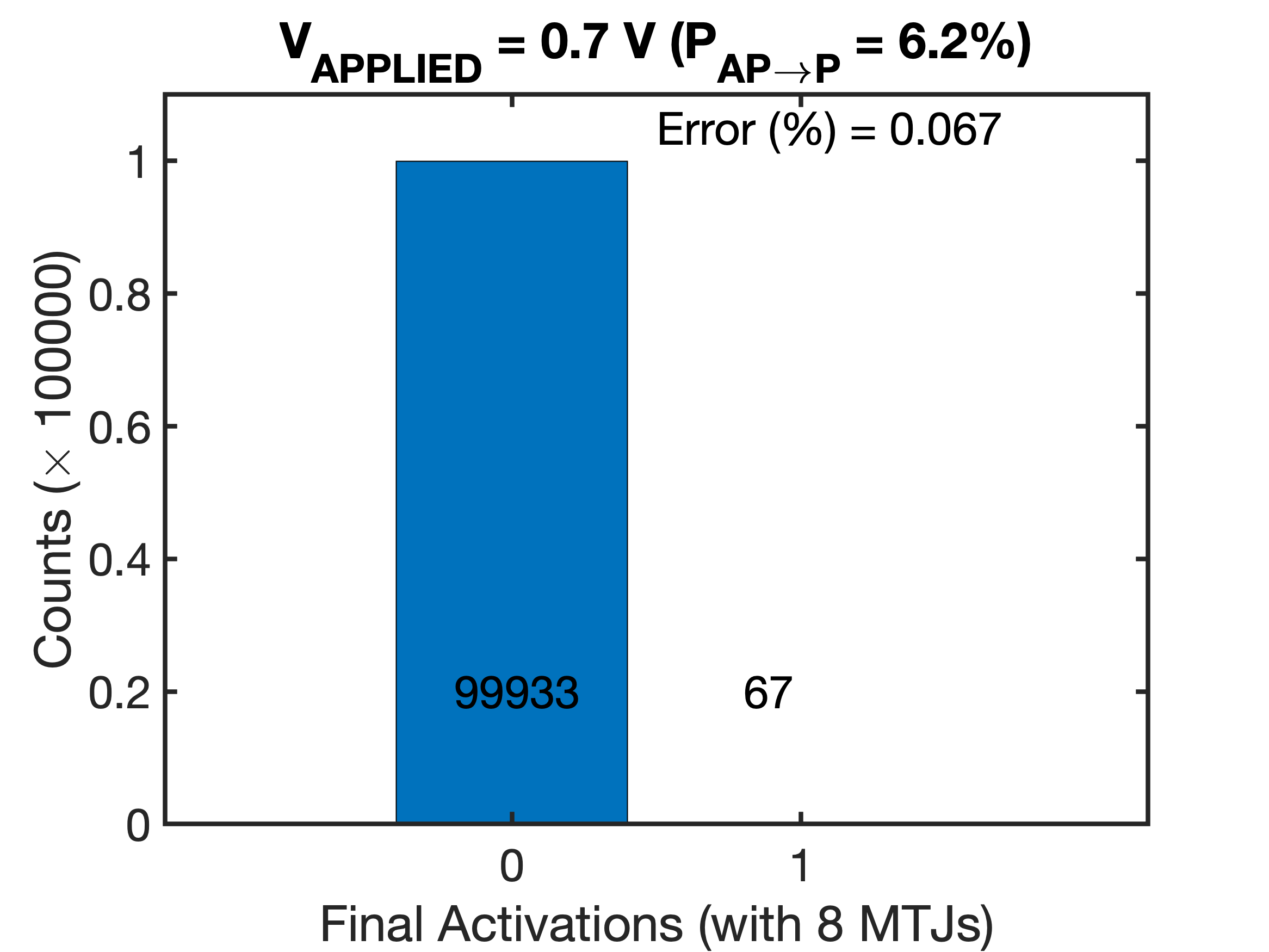}} \\
\subfloat[]{\includegraphics[width=0.5\linewidth]{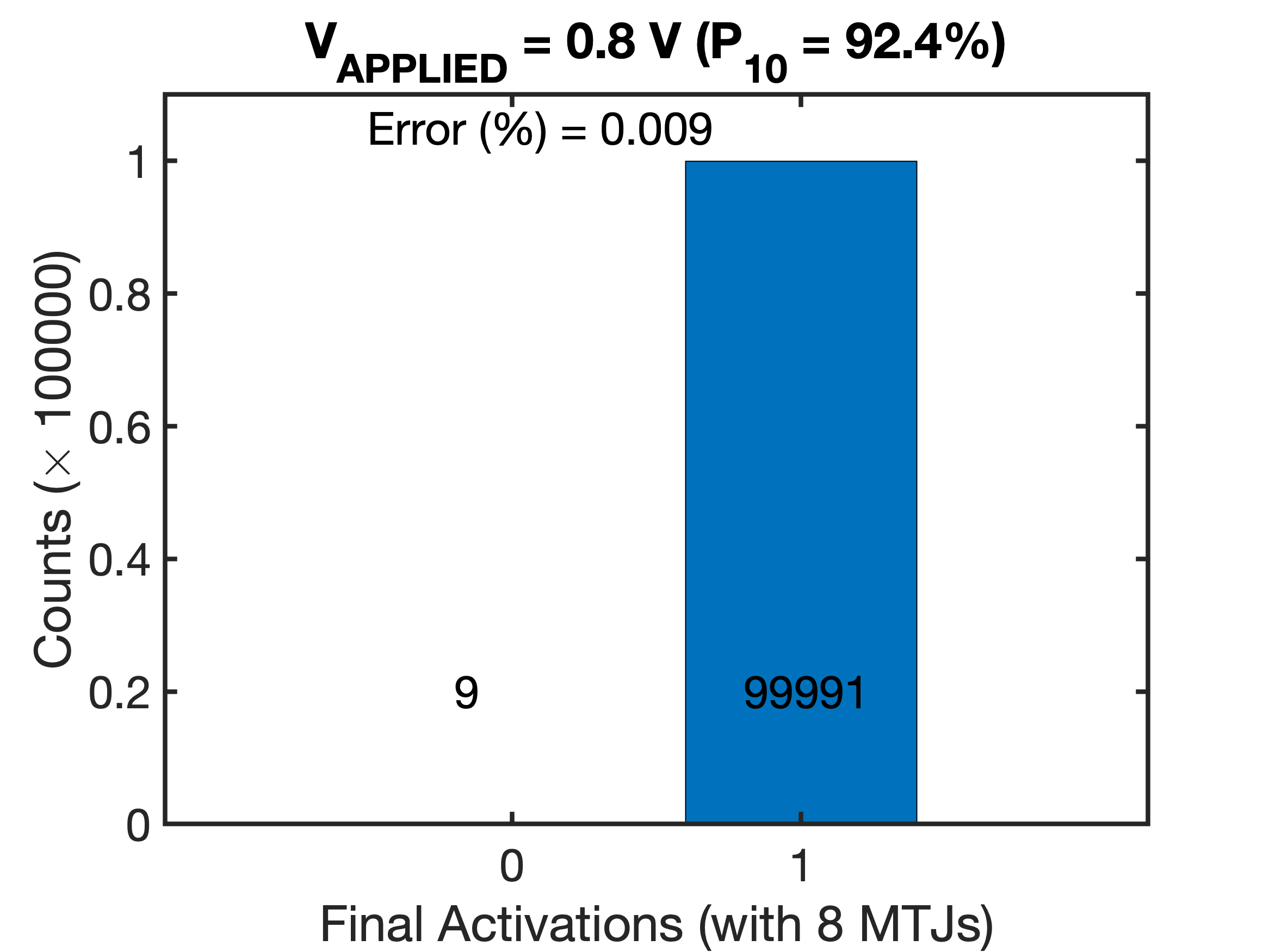}}
\subfloat[]{\includegraphics[width=0.5\linewidth]{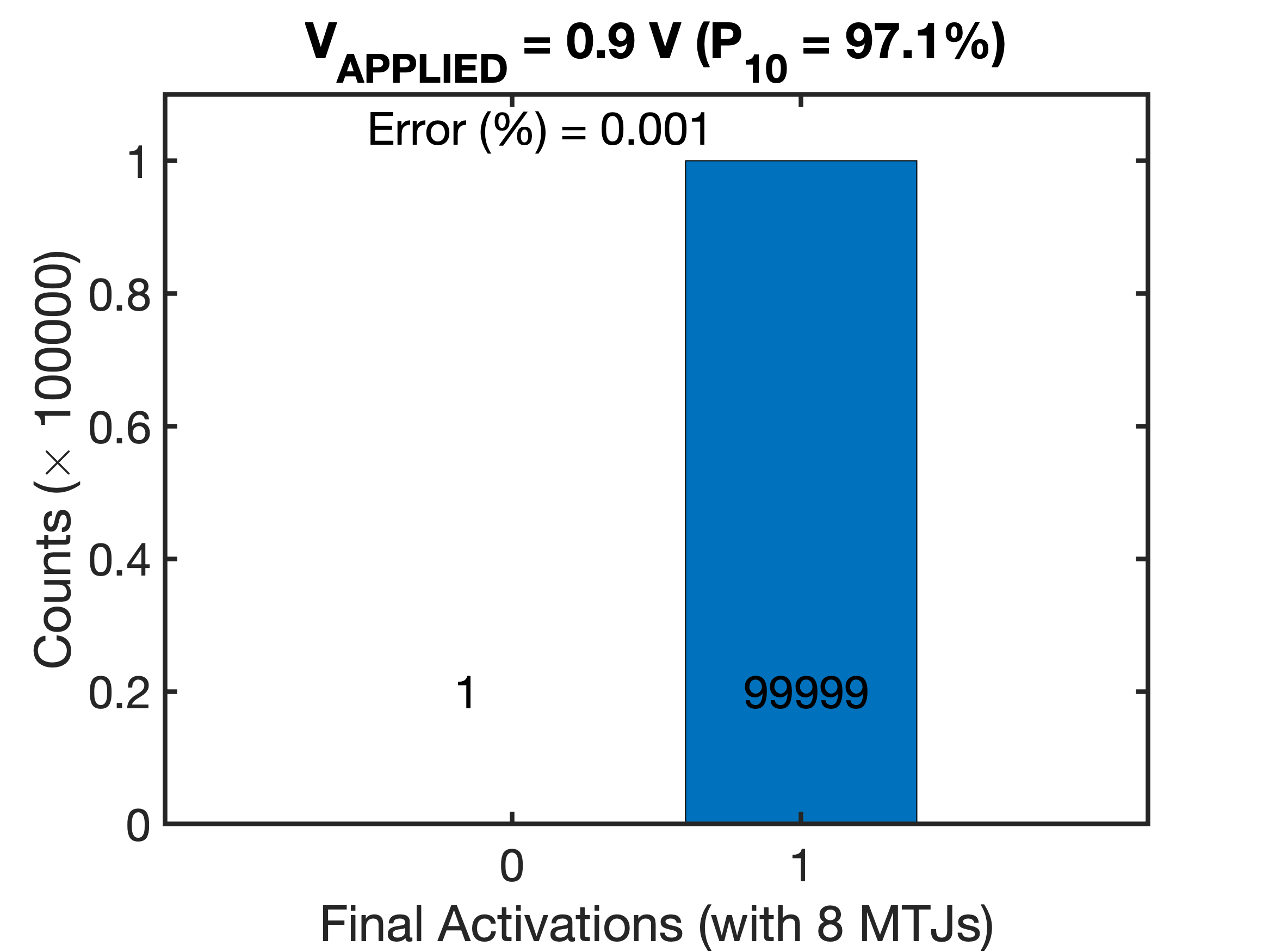}}
\caption{Final activation analysis for multiple VC-MTJs at various experimentally measured single VC-MTJ's anti-parallel to parallel switching probabilities of (a) 6.2\%, (b) 92.4\%, and (c) 97.17\%, respectively. By employing majority threshold operation with multiple VC-MTJs, the error rate for non-switching at 0.7 V and switching at 0.8V and 0.9V decreases to below 0.1\%. Note, we utilize 0.8V threshold in our hardware which ensures high-confident binary activation for our neural network.}
\label{redundant_prob}
\end{figure}

Figure \ref{redundant_prob} displays the reduction in error percentage achieved by employing multi-VC-MTJs to estimate the final activations. It can be seen that the error percentage decreases to less than 0.1\%, with individual VC-MTJ switching errors of 6.2\%, 7.6\%, and 2.9\% observed experimentally for the applied voltages of 0.7V, 0.8V, and 0.9V, respectively. The reduction in error percentage is due to majority operation implemented by effectively generating an output binary activation only if majority of VC-MTJs switch in a set of 8 VC-MTJs per kernel.

\subsubsection{Burst-mode Global Shutter Read and Reset Operation}
We employ the non-volatile VC-MTJ to generate and store the output activation signals per kernel, mitigating motion blur effects induced by the rolling shutter phenomenon (sequential exposure of rows instead of simultaneous exposure). During the read phase, we perform sequential reading of all the VC-MTJs in the pixel array followed by a reset operation when necessary. This effectively transforms energy and latency expensive ADC reads in a conventional camera into burst memory read operations, as the VC-MTJs are inherently non-volatile memory devices. Utilizing VCMA effect enables disturb-free and fast reading using the VC-MTJ. Due to the narrow sense margin, we opt for sequential reading instead of reading all MTJs in parallel. Additionally, a reset is needed for individual VC-MTJs based on its resistance state for subsequent evaluations. Hence, we initiate the read phase first to generate the output activation, followed by activating the reset pulse if the specific VC-MTJ is found to have been switched based on the accumulated analog convolutional voltage. In this study, we employ the anti-parallel state of the VC-MTJ as the reset state, and an appropriate pulse (0.9V, 500 ps) can transition the VC-MTJ resistance state to the anti-parallel from the parallel state. Iterative reset can be used to ensure deterministic switching. In the read phase, if the VC-MTJ generates a spike (parallel resistance state), the comparator (shown in Fig. \ref{circuit}(g)) outputs VDD (binary activation), or if the VC-MTJ is in the anti-parallel state, the sense amplifier remains at 0. The sequential read phases are controlled by appropriate read pulses at the MUX block (illustrated in Fig. \ref{circuit}(f)).

\begin{figure}[!t]
\centering
\includegraphics[width=0.75\linewidth]{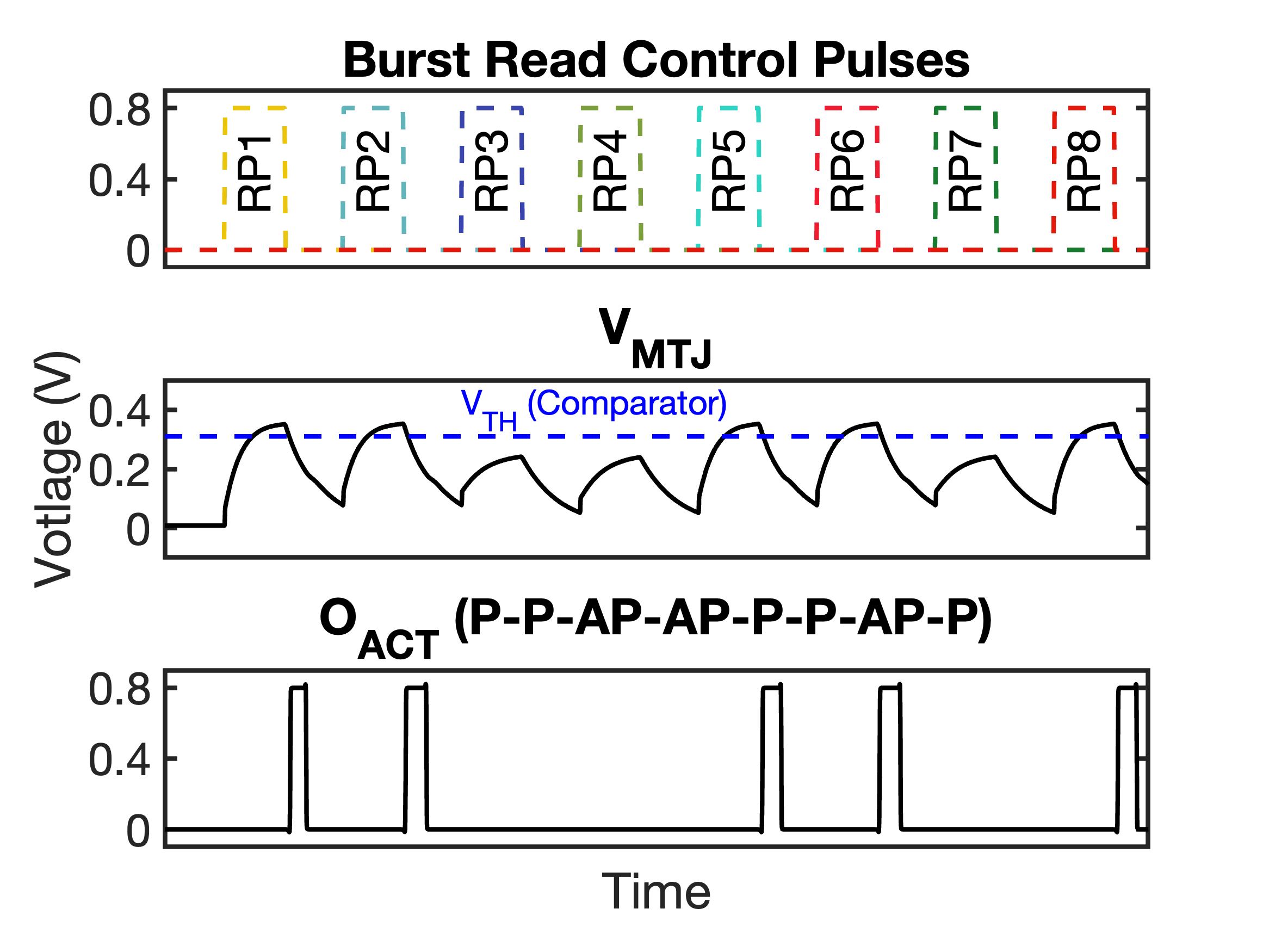}
\caption{Control pulses to enable burst-mode read for multi-VC-MTJs sequentially, the comparator input voltages (\si{V_{MTJ}}) along with the comparator threshold (blue dotted line) and output activation (\si{O_{ACT}}) for neuron states of P-P-AP-AP-P-P-AP-P, where P and AP denotes the parallel state (activated) and anti-parallel state (not activated) of the multi-VC-MTJ neurons, respectively.}
\label{seq_read}
\end{figure}

Figure \ref{seq_read} illustrates the burst-read operation of the VC-MTJ neurons implemented in our hardware. During the read phase, the source line (SL) is connected to the read voltage, which is set below the threshold for VC-MTJ switching to prevent unintended writes during reading. Additionally, the VCMA effect (discussed in the subsection \ref{device_details}) ensure disturbance-free reading. We sequentially compare the resistance state of the VC-MTJ neurons by activating the MUX switches, as depicted in the top subplot of Figure \ref{seq_read}. When a VC-MTJ neuron is in the parallel state, the voltage (\si{V_{MTJ}} exceeds comparator's threshold, thereby generating a spike. In the depicted scenario, 8 sequential burst pulses read 8 VC-MTJs in P, P, AP, AP, P, P, AP, P states, respectively. With 5 out of 8 MTJs switching from the reset state (AP), 5 output activation pulses (\si{O_{ACT}}) are generated, as shown in the bottom subplot of Figure \ref{seq_read}.

\subsection{Implementation of Binary Activation Neural Network Algorithm}\label{algo_details}

Our proposed BNN amenable to the hardware implementation described above, taking into account the binary switching characteristics of VC-MTJ devices, is formulated as 
\begin{equation}
    \boldsymbol z_l = \frac{\boldsymbol u_l}{v_l^{th}} \; \; \; \; \
    \boldsymbol{o}_l = \begin{cases} 1, & \text{if } \boldsymbol{z}_l \geq 1; \\ 
                                      0, & \text{otherwise} \end{cases} \label{eq:lif_1step}
\end{equation}
where $\boldsymbol{z}_l$ denotes the normalized activation output, $v_l^{th}$ denotes a trainable threshold parameter, and $\boldsymbol u_{l}\in\mathbb{R}^{c\times{h'}\times{w'}}$ is the output of the convolution operation in layer $l$, given input $\boldsymbol o_{l{-}1}\in\mathbb{R}^{c\times{h}\times{w}}$ and weight $\boldsymbol w\in\mathbb{R}^{n\times{c}\times{k}\times{k}}$. Note that such a sparse BNN is an excellent candidate for our processing-in-pixel architecture as it not only reduces the bit-precision of the output compared to the 8-bit unsigned raw pixel values but also yields lots of \textit{zeros} which can significantly reduce the communication energy via sparse coding schemes. Both these factors help reduce the bandwidth of our processing-in-pixel system. This sparse BNN is also similar to spiking neural networks (SNN) \cite{neuro_frontiers,spike_ratecoding,dsnn_conversion1} without the temporal dimension, i.e., only with one time step, which similarly yields significant activation sparsity. There are also one-time-step SNNs proposed in the literature \cite{one-time-step_SNN} that have the same network architecture as our sparse BNN.

However, optimizing a binary neuron with a unit step activation function as shown in Eq. \ref{eq:lif_1step} proves challenging, even with the straight through estimator (STE) based approaches \cite{bengio2013estimating} that either approximate the binary neuron's functionality with a continuous differentiable model or resort to surrogate gradient approaches used in SNNs \cite{lee_dsnn}. This difficulty arises because the activations predominantly contain zeros, hindering weight adjustments through gradient descent. In particular, if a pre-synaptic neuron outputs a zero, the connected synaptic weight cannot be updated, because its gradient is calculated as the product of the activation's gradient and the output of the pre-synaptic neuron. Consequently, it becomes crucial to lower the threshold value to generate more \textit{ones} in the activations, facilitating better network convergence. However, excessively low thresholds may result in each neuron producing lots of \textit{ones}, leading to random outputs in the final classifier layer. Thus, determining the optimal threshold for each layer poses a significant challenge.

To mitigate this concern, we leverage recently proposed Hoyer regularized spiking neural networks \cite{datta2024can} that dynamically down-scales the threshold based on the activation using Hoyer regularized training. Training the sparse BNN with a Hoyer regularizer can shift pre-activation distributions
away from the Hoyer extremum in a deep NN. Our principal insight is that setting our activation threshold to this extremum shifts the distribution away from the threshold value, reducing
noise and improving convergence \cite{datta2024can}. Specifically, we clip the activation of each convolutional layer to the trainable threshold obtained from gradient descent with our Hoyer loss function, as elaborated later in Equation \ref{eq:Hoyer_spike_func}. Thus, the normalized down-scaled threshold value, against which we compare the normalized activation for each layer, is computed as the Hoyer extremum of the clipped activation tensor.

\begin{equation}
    \boldsymbol{z}_l^{clip}{=}\begin{cases} 1,  \text{if } \boldsymbol{z}_l{>}1 \\ \boldsymbol{z}_l, \text{if } 0{\leq}\boldsymbol{z}_l{\leq}1  \\ 0, \text{if } \boldsymbol{z}_l < 0
    \end{cases} \ \ \
    \boldsymbol{o}_l{=} h_s(\boldsymbol{z}_l){=}  \begin{cases} 
    1, \text{if } \boldsymbol{z}_l{\geq}E( {\boldsymbol{z}_l^{clip}}) \\ 
    0, \text{otherwise} \end{cases}
    \label{eq:Hoyer_spike_func}
\end{equation}
\noindent

Note that our normalized threshold $E( {\boldsymbol{z}_l^{clip}})$ is less than the normalized threshold whose value is $1$ for any output. Hence, our actual threshold value $E( {\boldsymbol{z}_l^{clip}}){\times}v_l^{th}$ is less than the trainable threshold $v_l^{th}$ used in earlier SNN works \citep{datta_date,rathi2020dietsnn}. This leads to more weight updates, yielding better convergence. For more details on the formulation of the Hoyer regularized loss function and the weight updates, please refer to \cite{datta2024can}.

\subsection{Device-Hardware-Algorithm Co-design:}\label{codesign_details} 

\begin{figure}[!t]
\centering
\includegraphics[width=0.75\linewidth]{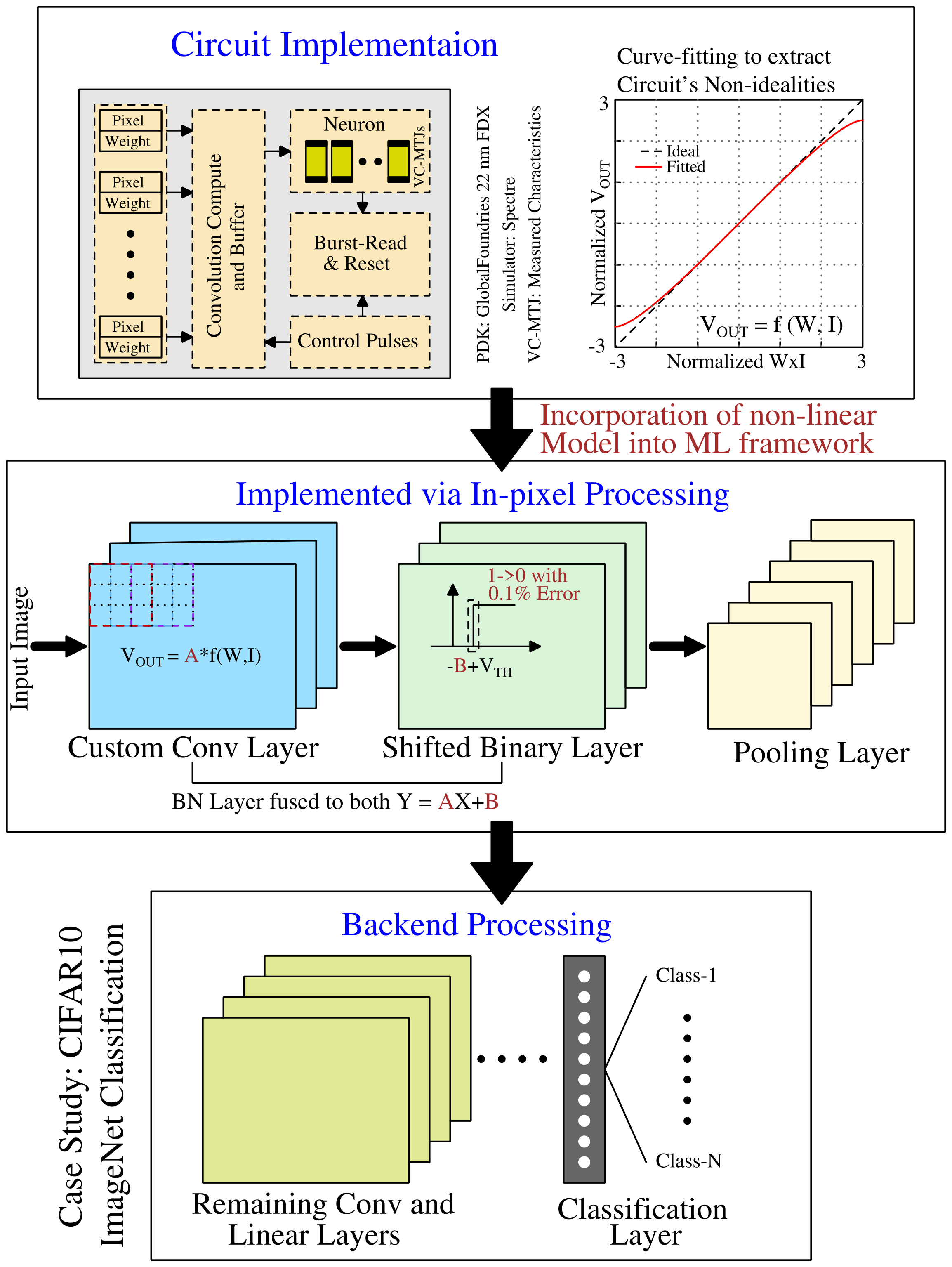}
\caption{Proposed device-circuit-algorithm co-design framework implementing our processing-in-pixel architecture.}
\label{codesign_image}
\end{figure}

\subsubsection{Custom Convolution Function of the First Layer representing Circuit's Non-idealities}

To simulate our in-pixel convolution accurately, we activate multiple pixels simultaneously as shown in Fig. \ref{scatter_plot_sub}(a). This allows the weight-modulated outputs of different pixels to be summed together in parallel in the analog domain, effectively emulating a convolution operation as discussed in \ref{inpix_ckt}. 

We incorporate this non-linear operation as a curve-fitted function incorporated in the high-level machine learning framework (PyTorch) adopted to this work as shown in Fig. \ref{codesign_image}. We also fuse the batch normalization layer by integrating the scale term into the preceding convolutional layer weights by embedding it directly into the pixel values of the weight tensor. Furthermore, we adjust the switching point of the MTJ-based comparator (which is ideally \si{V_{TH}}), as illustrated in Figure \ref{codesign_image}, to include the shift term B. This hardware-aware algorithmic approach ensures capturing the accurate hardware behavior while evaluating the system's performance. 

\subsubsection{Adjustable Threshold Matching of the VC-MTJs}
The near-deterministic switching threshold of the VC-MTJ depends on the device geometry. Therefore, to employ the VC-MTJ as a neuron for enabling global shutter in-pixel operation in our system, we repurpose our analog subtractor block to achieve required threshold and tunability, as detailed in the subsection \ref{inpix_ckt}. This device-hardware co-design approach allows us to accommodate the algorithmic requirement for variable thresholds and tunability.

\subsubsection{Use of Multiple VC-MTJs}
\begin{figure}[!t]
\centering
\includegraphics[width=0.75\linewidth]{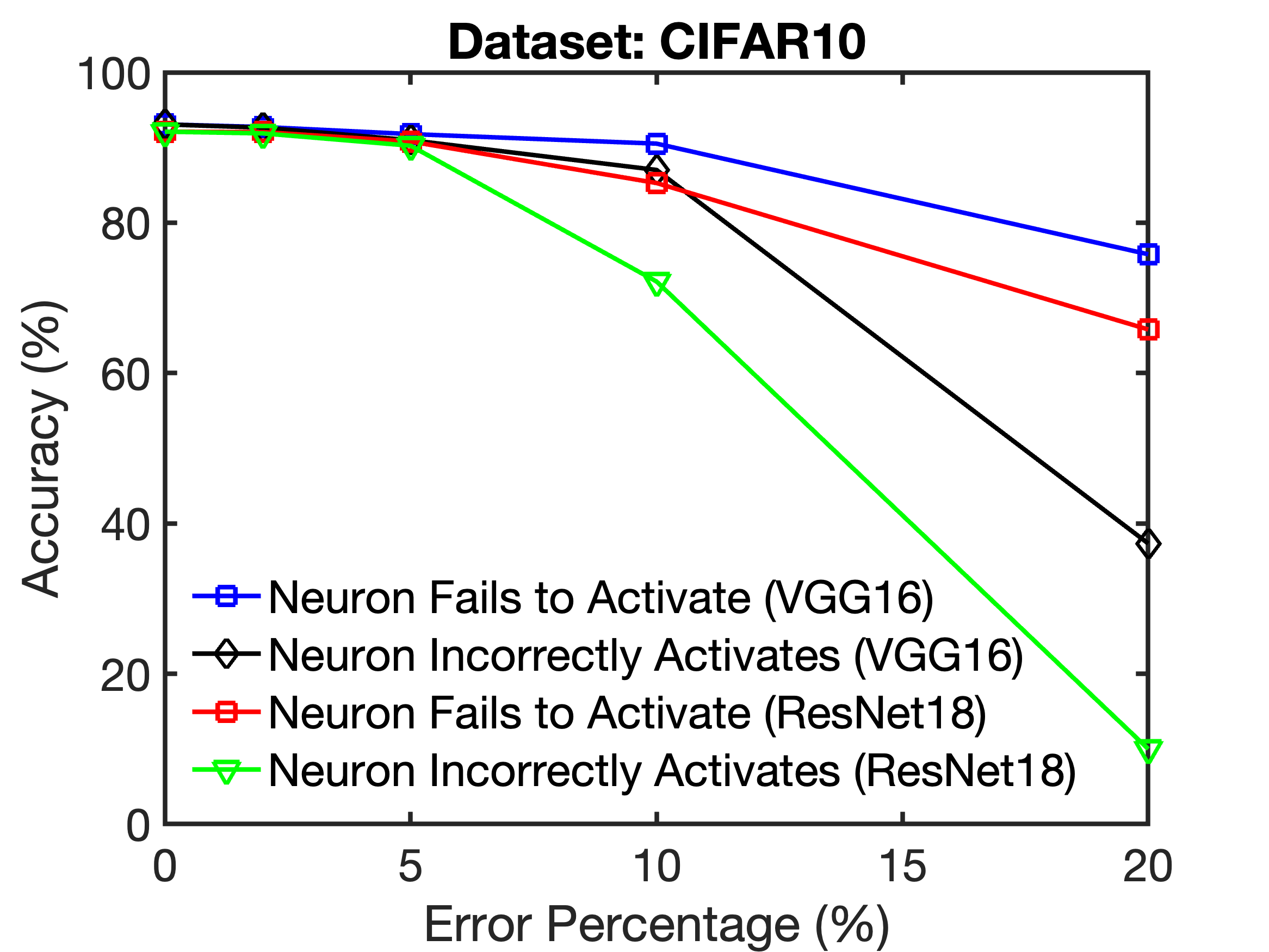}
\caption{Impact of the binary activation error percentage on the test accuracy with VGG16 and ResNet18 architectures on CIFAR10. Here, 'neuron fails to activate' (`neuron incorrectly activates') indicates that despite the convolution results exceeding (not exceeding) the binary threshold, the neuron does not activate (activates) due to its non-deterministic switching characteristics.}
\label{algo_error_prob}
\end{figure}

The near-deterministic nature of the MTJ neuron causes the sparse binary output of the first convolution layer implemented in the sensor to switch from 0 to 1 and vice-versa. This stochastic switching significantly degrades the accuracy of the BNN as shown in Fig. \ref{algo_error_prob}, particularly when the switching error is more than 10\% for \si{1\rightarrow0} transition and 3\% for \si{0\rightarrow1} transition. To support the high-confident switching needs of the algorithm for achieving better accuracy, we introduce multiple VC-MTJs as neuron (shown in Fig. \ref{circuit}(e)). The analysis exhibits that by utilizing 8 MTJs, the switching error can be reduced to less than 0.1\% as shown in Fig. \ref{redundant_prob}. Though our fabricated VC-MTJs exhibit near-deterministic switching (higher than 92\%) they still fall short of the algorithmic requirement for high confidence switching. Employing the multiple VC-MTJs, we circumvent high-confidence switching requirement of the algorithm to achieve accuracies close to state-of-the-art (SOTA) as shown later in section \ref{results_acc}.

\subsubsection{Network Optimization} 
We binarize the in-pixel output i.e., limit the output bit-precision to $1$ due binary switching nature of the VC-MTJs. The potential accuracy drop due to this extreme precision reduction is mitigated by the Hoyer regularized training method as illustrated in Section \ref{algo_details}. Additionally, to honor the area footprint limitations of the in-pixel hardware, for example, the typical pixel pitch limitations for fabrication, we limit the number of channels in the first convolution layer implemented in the pixel. In particular, we use $32$ channels, instead of $64$ channels that are common in both VGG and ResNet architectures in the first layer. We employ the standard stride of $2$, as reducing the stride to $1$, which results in non-overlapping kernel sliding, leads to significant drop in accuracy.

\begin{table*}
\caption{Comparison of the test accuracy of our BNNs with iso-weight-precision DNNs for object recognition. Model$^{*}$ indicates that we remove the first max pooling layer, and Sp. denotes sparsity. We set the switching probability of both \si{0\rightarrow1} and \si{1\rightarrow0} transition to $0.1\%$ as per our hardware demonstration.}
\label{tab: Results in object recognition}
\begin{center}
\centering
\setlength{\tabcolsep}{3.4mm}{
\begin{tabular}{l|c|c|c|c}
\hline
Network & dataset & DNN (\%) & BNN (\%) & Sp. (\%)\\
\hline
\hline
VGG16 & CIFAR10 & 94.10  & 93.08 &   79.24 \\
\hline
ResNet18 & CIFAR10 & 93.34  & {92.11} &  72.59  \\
\hline
ResNet18$^{*}$ & CIFAR10 & 94.28  & {93.46} & 82.59 \\
\hline
ResNet20 & CIFAR10 & 93.18 & {92.24} &  76.50 \\
\hline
ResNet34$^{*}$ & CIFAR10 & 94.68  & {93.40} & 83.29 \\
\hline
ResNet50$^{*}$ & CIFAR10 & 94.90  & {93.71} & 83.54\\
\hline
\hline
VGG16 & ImageNet & 70.08  & 67.72 &  75.22 \\
\hline
\end{tabular}
}
\end{center}
\end{table*}

\section{Results and Discussion}\label{results} 

\subsection{Accuracy}\label{results_acc}
We conduct object recognition experiments on CIFAR10 \cite{cifar} and ImageNet \cite{imagenet_cvpr09} dataset using VGG16 \cite{vgg} and several variants of ResNet \cite{resnet} architectures. For training these models, we use the Adam \cite{kingma2014adam} optimizer for VGG16, and use SGD optimizer for ResNet models. As shown in Table \ref{tab: Results in object recognition} {with 4-bit weights}, we obtain the close to SOTA accuracy of $93.08\%$ on CIFAR10 with VGG16; the accuracy of our ResNet-based sparse BNNs are also close to the iso-weight-precision DNN counterparts. On ImageNet, we obtain a $67.72\%$ top-1 accuracy with VGG16 which is only ${\sim}2.3\%$ lower compared to the iso-architecture full-precision counterpart. 

\subsection{Bandwidth}\label{subsec:BW}

The bandwidth reduction from our processing-in-pixel hardware can be estimated as 

\begin{equation}\label{eq:pip_compression}
    C = \left(\frac{{h^{out}}{\times}{w^{out}}{\times}{c^{out}}}{{h^{in}{\times}{w^{in}}{\times}{c^{in}}}}\right)\cdot\frac{b_{inp}}{b_{out}}\cdot\frac{4}{3}
\end{equation}

where $h$, $w$, $c$ represent the height, width, and the number  of channels respectively and the superscript ${inp}$ and ${out}$ represent the image input and in-sensor output (after convolutional, batch norm, binary, and pooling layers) respectively. $b_{inp}$ denotes the pixel bit-precision (${\sim}12$ in image sensors, and $b_{out}{=}1$ denotes the in-sensor output bit-precision. The factor $\left(\frac{4}{3}\right)$ represents the compression from Bayer’s pattern of RGGB pixels to RGB pixels. 
Plugging these values for VGG16, we obtain $C{=}6$. All our sparse models yield a sparsity of ${\sim}75\%$ or higher in the output of the in-sensor layer on both CIFAR10 and ImageNet, which is significantly higher compared to existing SNNs/BNNs. This provides the opportunity to further reduce the bandwidth (even more than $6\times$) via effective sparse coding schemes, such as compressed sparse row/column based communication.

\begin{figure}[!t]
\centering
\includegraphics[width=0.75\linewidth]{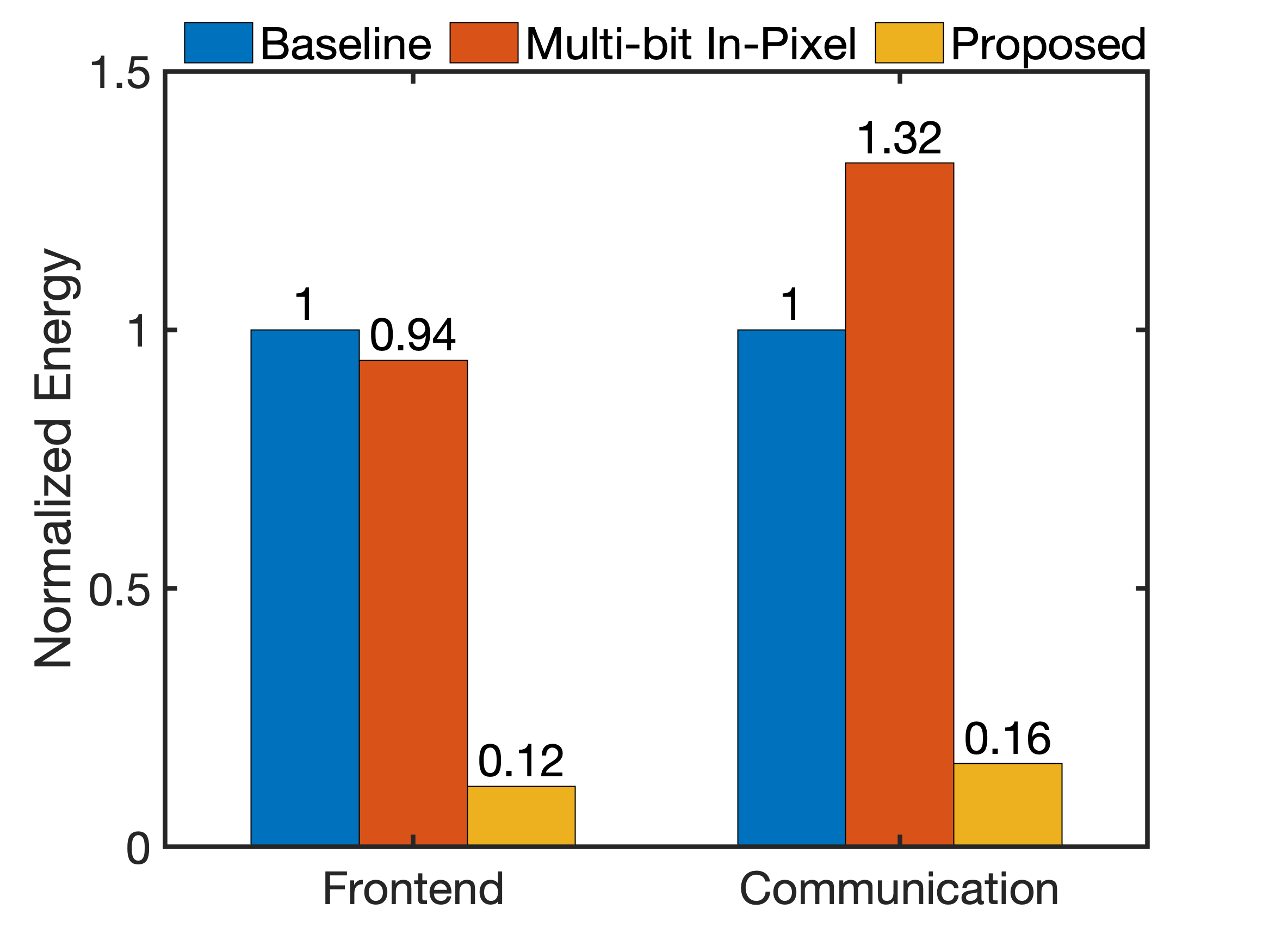}
\caption{Comparison of the normalized front-end and communication energy from the sensor to the back-end hardware of our proposed processing-in-pixel architecture with the state-of-the-art in-sensor computing architecture \cite{aps_p2m} and a baseline architecture where the entire network is implemented outside the sensor, for processing the ImageNet dataset with VGG16 architecture. Note, the energy is normalized with respect to the baseline energy.}
\label{energy_plot}
\end{figure}

\subsection{Energy Efficiency}

We compare the energy incurred by our processing-in-pixel architecture with the state-of-the-art in-sensor computing \cite{aps_p2m} and a baseline architecture where the entire BNN is implemented outside the sensor (where the sensor only reads out the analog pixel values, converts them to digital outputs using ADCs, and sends them to the back-end hardware processing the network) in Fig. \ref{energy_plot}. As we can see, our solution reduces the front-end energy i.e., the energy consumed by the sensor, compared to the existing in-sensor and baseline counterparts by $8.0{\times}$ and $8.2{\times}$ respectively. This is due to our efficient MAC and comparison operation in the pixels, and most importantly, the removal of ADCs due to binary nature of the pixel output, that otherwise dominate the sensor energy. The binary nature also reduces the communication energy (from the sensor to back-end hardware) by up to $8.5{\times}$ compared to the other approaches as shown in Fig. \ref{energy_plot}. We have estimated the energy number from the simulation results obtained utilizing GlobalFoundries 22nm FDX technology node considering an integration time of 5 \si{\mu s}, write and read pulse widths per VC-MTJ are of 700 ps and 500 ps, respectively. Note that we ignore the back-end energy calculation since that depends on the underlying hardware architecture and dataflow which is not the focus of this work.

It is important to note that the communication energy is calculated based on the low voltage differential signaling (LVDS) data link under the premise that the front-end and back-end hardware are situated in close proximity on the same printed-circuit board. However, in practice, front-end and back-end sensors may be separated by significant distances, necessitating long and energy-intensive wired or wireless data transfers \cite{aps_p2m_detrack}. This scenario is common in applications such as sensor fusion and swarm intelligence. In such instances, the overall energy savings would tend to align with the bandwidth reduction, which can be as high as $6\times$, as shown in the subsection \ref{subsec:BW}.

\subsection{Latency}
Thanks to the localized in-pixel compute unit positioned near the pixel array and the utilization of non-volatile VC-MTJ, we achieve high-speed and global shutter functionality. The absence of the need to charge the large bitline capacitance per pixel, owing to the proximity of the compute unit, allows for very short integration times, such as 5 \si{\mu s}. Our proposed in-pixel approach enables simultaneous storage of all activations in the VC-MTJ neurons. Additionally, our fabricated VC-MTJ demonstrates sub-nanosecond disturb-free read characteristics, enabling burst memory read operations instead of the conventional analog-to-digital conversion found in traditional camera sensors. All computations can be completed utilizing two integration times—one for negative weights and one for positive weights, as outlined in subsection \ref{inpix_ckt}—followed by a burst memory read operation of all the neurons. The convolution computation and reading of all neurons can be completed in under 70 \si{\mu s}, assuming an image size of 224\si{\times}224, a kernel size of 3\si{\times}3\si{\times}3, and a stride of 2, with reads happening in sub-nanosecond time frames.

\section{Conclusions}\label{conclusion}
We present an energy and bandwidth efficient in-pixel processing solution for edge devices targeted computer vision applications. Leveraging the state-of-the-art nanoscale high endurance, high-speed, and non-volatile VC-MTJ, we enable global shutter in-pixel computation. Additionally, utilizing optimized binary activation neural networks to compute feature spikes efficiently, our ADC-less architecture exhibits energy and bandwidth improvement compared to the traditional systems. Our device-circuit-algorithm co-design framework incorporates device and hardware constraints, facilitating accurate evaluation of system performance. Through extensive evaluations on CIFAR10 and ImageNet datasets, we demonstrate significant improvements in energy efficiency and bandwidth reduction compared to traditional systems, without compromising test accuracy. These contributions advance the field of energy-efficient edge processing and pave the way for future innovations in extreme-edge intelligence.

\bmhead{Acknowledgements}
The research was funded in part by National Science Foundation through awards CCF2319617 and CCFCCF-2322572 and Intel Neuromorphic Computing Lab.

\section*{Authors Contribution Statement}
M.K. conceptualized the idea, designed and verified the circuits, and developed the device-circuit co-design framework. G.D. planned and implemented the Hoyer regularized training, carried out the algorithmic simulations and developed the device-circuit co-design framework. J.A. conducted the experimental design, data collection, and analysis for the characterization and switching behavior of magnetic tunnel junctions (MTJs). C.D. assisted with data collection and analysis for the characterization and switching behavior of the MTJ. A.J.$^1$ contributed to the proposed integration scheme of CIS and MTJs. P.A. and P.B. supervised the overall research. A.J.$^2$ also conceptualized the idea and supervised the overall research. All authors contributed to reviewing the manuscript. Note that AJ$^1$ and AJ$^2$ are Ajey P. Jacob and Akhilesh R. Jaiswal respectively.


\bibliography{sn-bibliography}

\end{document}